# Electrically active defects in 3C, 4H and 6H silicon carbide polytypes: A review

Ivana Capan

Ruđer Bošković Institute, Bijenička 54, 10 000 Zagreb, Croatia

**Abstract:** This paper aims to critically review electrically active defects studied by junction spectroscopy techniques (deep-level transient spectroscopy and minority carrier transient spectroscopy) in the three most commonly used silicon carbide (SiC) polytypes: 3C-SiC, 4H-SiC, and 6H-SiC. Given the dominant role of SiC in power electronic devices, the focus is strictly on electrically active defects that influence material performance and device reliability. The most prevalent defects in each polytype and their effects on electrical properties will be examined. Additionally, recent advancements in defect characterization and defect engineering will be highlighted, emphasizing their impact on improving SiC-based device performance. The paper will also address the main challenges that continue to hinder the broader adoption of SiC, such as defect-related limitations in carrier lifetime and doping efficiency. Furthermore, beyond the well-established applications of SiC in power electronics and high-temperature environments, lesser-known niche applications will be explored, showcasing the material's versatility in emerging fields.

## 1. Introduction

Semiconductors are the pillars of modern electronics, a driving force bringing innovations in different industries, from consumer electronics to renewable energy systems. As global demand for energy-efficient and high-performance devices continuously grows, traditional silicon-based devices face limitations in handling the increasingly stringent requirements of high-power, high-temperature, and high-frequency applications. These challenges have spurred widespread interest in alternative semiconductor materials, among which silicon carbide (SiC) has emerged as a game-changer. Silicon carbide is a wide bandgap semiconductor with unique material properties, such as a large critical electric field, high thermal conductivity, high electron saturation velocity, chemical inertness, and radiation hardness [1]. These properties make SiC-based devices far superior to their silicon counterparts for demanding applications. Furthermore, SiC has already established itself as a critical enabler of advancements in power electronics, where its capabilities have driven breakthroughs in efficiency and performance [2-4]. Recent progress in producing high-quality SiC wafers has opened the door to additional applications, including quantum technologies [5-7] and radiation detection [8-13], further solidifying SiC's pivotal role in next-generation semiconductor technology.

The most commonly used SiC polytypes are 3C-, 4H-, and 6H-SiC. Accordingly, this literature review focuses on these three polytypes. Table 1 summarizes the fundamental material properties of 3C, 4H, and 6H-SiC [14, 15].

**Table 1.** Material properties of SiC polytypes. Data are adapted from [14, 15].

| SiC polytype | Crystal structure | Energy band gap (eV) | Electron mobility ∥ / ⊥ to *c*-axis ($cm^2V^{-1}s^{-1}$) | Hole mobility ($cm^2V^{-1}s^{-1}$) | Thermal conductivity ($Wcm^{-1}K^{-1}$) | Electric field ∥ to c-axis (MV/cm) |
|---|---|---|---|---|---|---|
| **4H** | Hexagonal | 3.26 | 1200/1020 | 120 | ~4.9 | 2.8 |
| **6H** | Hexagonal | 3.02 | 100/450 | 100 | ~4.5 | 3.0 |
| **3C** | Cubic | 2.36 | ~1000/1000 | 100 | ~3–4 | 1.4 |

Despite the abundance of original research papers and numerous SiC-related literature reviews, focused review articles that provide a comparative overview of the 3C, 4H-, and 6H-SiC polytypes remain surprisingly scarce. Among the limited number of such reviews, a few stand out and are highly recommended:

(i) Deep Defect Centers in Silicon Carbide Monitored with Deep Level Transient Spectroscopy by Dalibor et al. [16]. This comprehensive review examined electrical data obtained from DLTS investigations on deep defect centers in various SiC polytypes. The study provided a comparative analysis of deep-level defects across different SiC polytypes, aiding in the broader understanding of defect physics in SiC materials.

(ii) Comparative Study on Silicon Carbide (SiC) Polytypes in High Voltage Devices by W.Taha [17]. This paper reviews the crystallography of major SiC polytypes and their electrical properties, presenting a material physical model and discussing their applications in high-voltage devices.

(iii) Electron Mobility in Bulk n-Doped SiC-Polytypes 3C-SiC, 4H-SiC, and 6H-SiC: A Comparison by C.G.Rodriguez [18]. This study presents a comparative analysis of charge transport in bulk n-type doped SiC polytypes, focusing on electron drift velocity and mobility under varying electric field intensities and orientations.

(iv) A Comparative Study of Schottky Barrier Heights and Charge Transport in SiC Polytypes by Mekaret et al. [19]. In this paper, a comparative analysis of Schottky diodes using three SiC polytypes (3C, 4H, and 6H), focusing on Schottky barrier heights and charge transport mechanisms.

(v) Comparison of 6H-SiC, 3C-SiC, and Si for power devices by Bhatagar et al. [20]. The authors compared the performance of power rectifiers and power metal-oxide-semiconductor field-effect transistors (MOSFETs) made from 3C- and 6H-SiC with those made of silicon. This study suggested that SiC power rectifiers and MOSFET's could be a superior alternative for all Si power devices with breakdown voltage as high as 5000 V.

A logical question arises: why is this review paper necessary? Unlike the aforementioned review articles, this paper focuses exclusively on recent advancements in understanding the electrically active defects in SiC obtained by junction spectroscopy techniques, making it an invaluable resource for the SiC research community for several key reasons. The first reason is to gather a significant amount of information therefore making it more accessible, for young researchers in particular. Research on different SiC polytypes and their electrically active defects is scattered across numerous studies. The second reason is to identify the differences between the electrically active defects in three polytypes and their implications for device applications. Hopefully, this will help researchers choose the most suitable polytype for specific applications. Finally, the third reason is to recognize and highlight problems that are still unresolved, steering future research efforts.

As this paper aims to provide a review of the main achievements, the Web of Science Core Collection was utilized to extract relevant data on SiC research publications up to the present day. Figure 1 illustrates the total number of publications associated with the keywords "3C-SiC," "4H-SiC," and "6H-SiC." Unsurprisingly, 4H-SiC-related papers dominate the dataset, as 4H-SiC has emerged as the preferred polytype for power electronics applications [2-4].

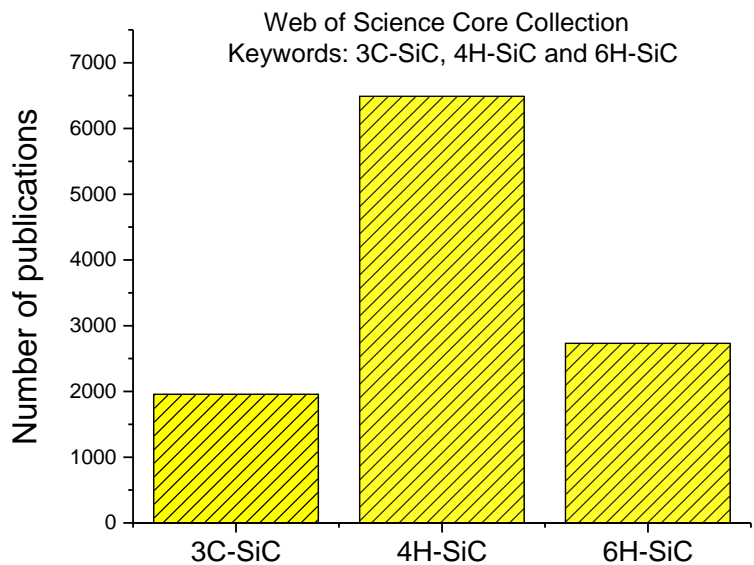

**Figure 1.** Total number of SiC publications. Data are extracted from the Web of Science Core Collection with keywords "3C-SiC", "4H-SiC" and "6H-SiC".

Given the predominant use of SiC in various electronic applications, it is essential to delve deeper into how these applications are distributed across the different polytypes. This analysis will provide valuable insights into the specific roles and advantages of each polytype in meeting the demands of modern electronic systems.

Using the Web of Science Core Collection and the three main SiC polytypes as primary keywords, additional keywords such as "diode" and "solar cell" were incorporated to refine the analysis. Interestingly, 3C-SiC emerges as the most represented polytype for solar cell applications, accounting for 64.8% of all related publications. In contrast, for a fundamental electronic device like a diode, the 4H-SiC polytype dominates, comprising 74.53% of the total publications. This distinction highlights the differing suitability and specialization of SiC polytypes for specific applications, driven by their unique material properties.

Electrically active defects play a critical role in all electronic device applications, as they act as traps for charge carriers and significantly influence the electrical properties of the material. These defects can lead to considerable deterioration in device performance and reliability. Electrically active defects may be introduced into the material at various stages, including during semiconductor growth, processing (e.g., ion implantation), or exposure to irradiation [14]. Deep-level transient spectroscopy (DLTS) has become indispensable for studying electrically active defects. DLTS is highly sensitive, capable of detecting electrically active defects at concentrations as low as $10^9$ cm$^{-3}$. Additionally, it provides critical information for determining a defect's "fingerprint," including activation energy for electron or hole emission, capture cross-section, and defect concentration [21]. The primary limitation of conventional deep-level transient spectroscopy (DLTS) lies in its insufficient energy resolution, making it nearly impossible to distinguish between closely spaced deep energy levels. This issue was significantly addressed by the development of Laplace DLTS (L-DLTS), an advanced variant of DLTS that offers an order of magnitude improvement in energy resolution, achieving precision in the millielectronvolt (meV) range [22]. While DLTS has been predominantly employed to study electrically active defects related to majority charge carrier traps, much less attention has been given to minority charge carrier traps. The foundational principles of minority carrier transient spectroscopy (MCTS) techniques were initially introduced by Hamilton et al. [23] and subsequently expanded upon by Brunwin et al. [24]. More details on transient spectroscopy techniques (DLTS, L-DLTS, MCTS) can be found in Ref.[21] and references therein.

This paper is organized as follows: Section 2 serves as a foundation by providing a comprehensive overview of electrically active defects in 4H-SiC, the most extensively studied SiC polytype due to its superior electronic properties and widespread use in power electronics. The section will cover key defects and their impact on device performance. Section 3 explores 6H-SiC, another important polytype, discussing its specific defect landscape and the implications for various applications. Section 4 shifts the focus to 3C-SiC, highlighting its unique advantages, such as its compatibility with silicon substrates and the most common electrically active defects that influence its performance. Section 5 presents a summary of the key findings. This section will also include a discussion on future research directions and emerging applications that could benefit from advancements in SiC defect engineering. By structuring the paper this way, the main aim is to provide a clear and detailed comparison of the three most widely used SiC polytypes, focusing on their electrically active defects and their influence on material performance and device applications.

## 2. 4H-SiC

This section provides an overview of the electrically active defects 4H-SiC. The focus is on intrinsic defects in n-type material introduced during the crystal growth and defects introduced during the irradiation. Due to the favorable material properties, n-type 4H-SiC is the preferred choice for most electronic applications. This has been reflected in the number of available studies in the literature, as the vast majority are devoted to n-type material. The p-type 4H-SiC is by far less studied compared to n-type 4H-SiC for several reasons, rooted in both material properties and application-driven priorities:

(i) Challenges in achieving high-quality p-type doping

Aluminum (Al) and boron (B) are commonly used dopants for creating p-type 4H-SiC. However, achieving high doping efficiency is difficult due to the high ionization energies of these acceptor impurities (~0.2 eV for Al and ~0.37 eV for B). These high activation energies result in a low fraction of active holes at room temperature, limiting the effectiveness of p-type doping [25].

(ii)   Power device applications are dominated by n-type 4H-SiC

In 4H-SiC-based power devices, such as Schottky diodes and MOSFETs, n-type 4H-SiC is primarily used because it serves as the main active region or drift layer. The superior electron mobility in n-type 4H-SiC compared to hole mobility in p-type 4H-SiC makes n-type material more suitable for high-power, high-frequency applications [1-4].

(iii)  Radiation detection applications prioritize n-type 4H-SiC

Research into radiation-induced defects has predominantly focused on n-type 4H-SiC [8-14]. Radiation detectors for alpha particles or neutrons are produced using the n-type material.

In summary, the less frequent study of p-type 4H-SiC is driven by the combination of its material challenges, inferior electrical properties, and its limited role in the primary applications of SiC-based devices. However, as new applications and technologies emerge, research on p-type 4H-SiC may gain more attention. p-type 4H-SiC has potential for applications in ultra-high-power electronics for high-voltage grids and high-speed trains [26].

We shall now focus on the most dominant and important electrically active defects in n-type 4H-SiC.

### 2.1 Carbon interstitials ($C_i$)

The study of electrically active radiation-induced defects in 4H-SiC has been an area of significant research for many years. A wide range of such defects, introduced by various radiation sources including protons, electrons, and neutrons, has been reported [27-33]. Among these, two electrically active defects with deep levels at 0.40 eV and 0.70 eV below the conduction band have garnered particular attention within the SiC research community. What makes these two defects especially intriguing is their consistent appearance together, regardless of the type of radiation source. For a long time, they remained enigmatic, and it was only recently that they were conclusively assigned to the same defect. The mystery surrounding their origin has finally been resolved. Storasta et al. [34] have used low-energy electrons to introduce only carbon displacements in the 4H-SiC lattice (C displacement threshold 20 eV, Si displacement threshold 35 eV). They have observed two energy levels at $E_C$ - 0.4 eV and $E_C$ - 0.7 eV, and labeled them as $EH_1$ and $EH_3$. These levels were assigned to a highly mobile defect, such as carbon interstitial. Alfieri et al. [35] provided compelling evidence that, under low-energy electron irradiation (E < 200 keV), only carbon interstitial-related defects are introduced, shedding light on the underlying mechanisms responsible for these two deep-level defects.

Figure 2 shows DLTS spectra for the low-energy electron (E =150keV) and fast neutron irradiated n-type 4H-SiC samples. In addition to $Z_{1/2}$ (which will be discussed in detail later), two deep-level defects at $E_c$−0.4 eV ($EH_1/S_1$) and $E_c$−0.7 eV ($EH_3/S_2$) are detected in irradiated 4H-SiC samples.

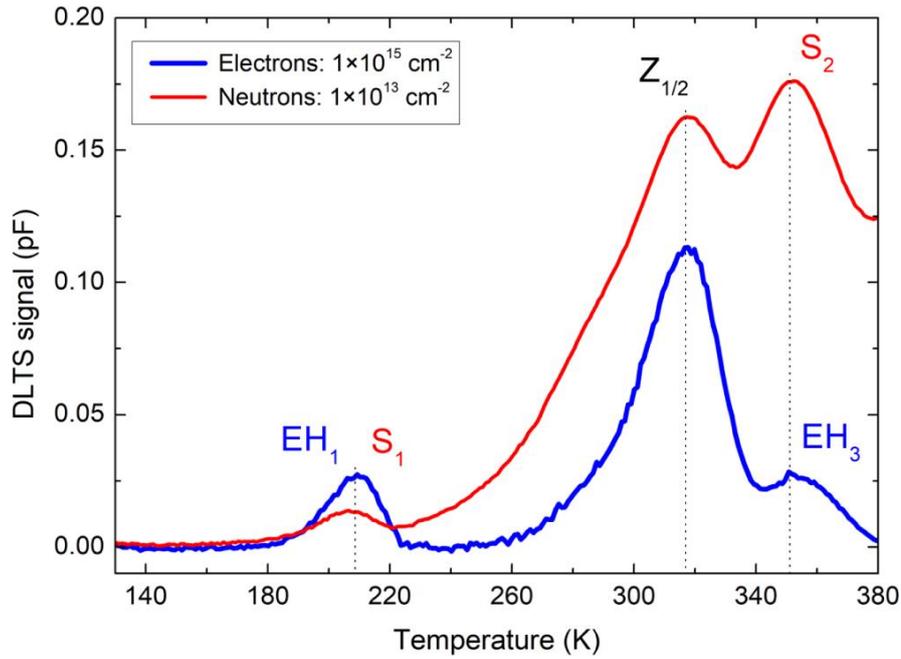

**Figure 2.** DLTS spectra for low-energy electron irradiated (blue) and neutron irradiated (red) n-type 4H-SiC samples. Data taken from Ref. [36].

2.2 Silicon vacancy ($V_{Si}$)

The $S_1$ and $S_2$ traps are introduced by high-energy particles such as ions, protons, and neutrons [29,32,33,36]. As seen in Figure 2, these traps exhibit the same energy levels as those of $EH_1$ and $EH_3$, located at $E_C$−0.40 eV and $E_C$−0.70 eV, respectively. This has caused significant confusion in distinguishing and identifying these defects over the years. Even today, when studying radiation-induced defects in SiC, it remains common practice within the research community to use the label "EH" to refer to a wide range of radiation-induced defects, further contributing to the ambiguity.

Bathen et al. [37] investigated the effects of MeV proton irradiation on 4H-SiC and observed, as expected, the two deep levels at $E_C$−0.40 eV and $E_C$−0.70 eV, designated as $S_1$ and $S_2$. These levels were attributed to the charge state transitions of silicon vacancies, specifically $V_{Si}$(−3/=) for $S_1$ and $V_{Si}$(=/−) for $S_2$. Using L-DLTS measurements, it was further revealed that the S1 level has two distinct emission lines corresponding to silicon vacancies located at the -*k* and -*h* lattice sites. These findings were later corroborated by Brodar et al. [38], who observed identical results when studying 4H-SiC materials irradiated with fast neutrons.

Moreover, Knežević et al. [36] have provided an additional approach to convincingly yet easily distinguish $C_i$ and $V_{Si}$. L-DLTS provides direct evidence that $EH_1$ consists of a single emission line arising from the $C_i$ at the −*h* lattice site (Figure 3a), in contrast to $S_1$ which has two emission lines arising from the $V_{Si}$ at the −*h* and −*k* lattice sites (Figure 3b).

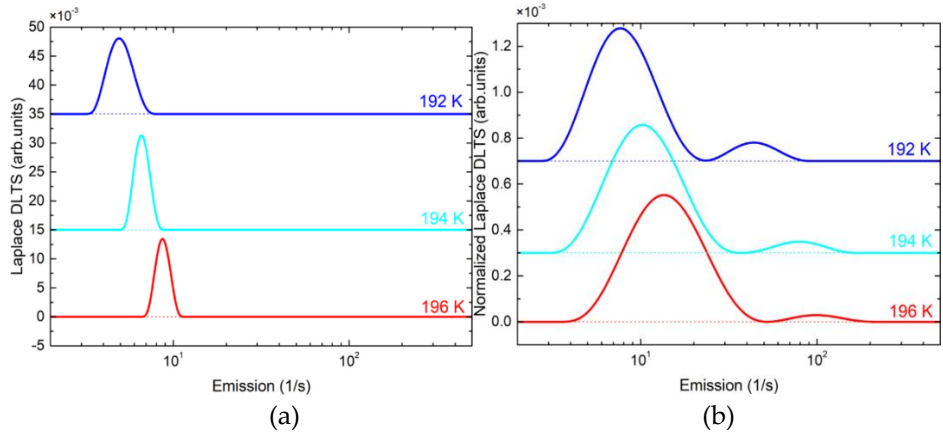

**Figure 3**. L-DLTS spectra for (a) $EH_1$, and (b) $S_1$ traps in low-energy electron and neutron irradiated n-type 4H-SiC, respectively. Data taken from Ref. [36].

Recently, $V_{Si}$ defects in 4H-SiC have gained significant attention within the SiC and quantum technology communities due to their unique properties, including half-integer spin, photostability, and long spin coherence times at room temperature [5,6,7,37]. The ability to efficiently introduce and control individual $V_{Si}$ defects in 4H-SiC is crucial for advancing quantum technologies. Several methods have been proposed, such as irradiation, ion implantation, and laser processing [39].

2.3 Carbon vacancy ($V_C$)

Figure 4 shows a typical DLTS spectrum for an as-grown n-type 4H-SiC sample. Two prominent traps, $Z_{1/2}$ and $EH_{6/7}$, are observed. The estimated activation energies for electron emissions are 0.67 eV and 1.64 eV, respectively.

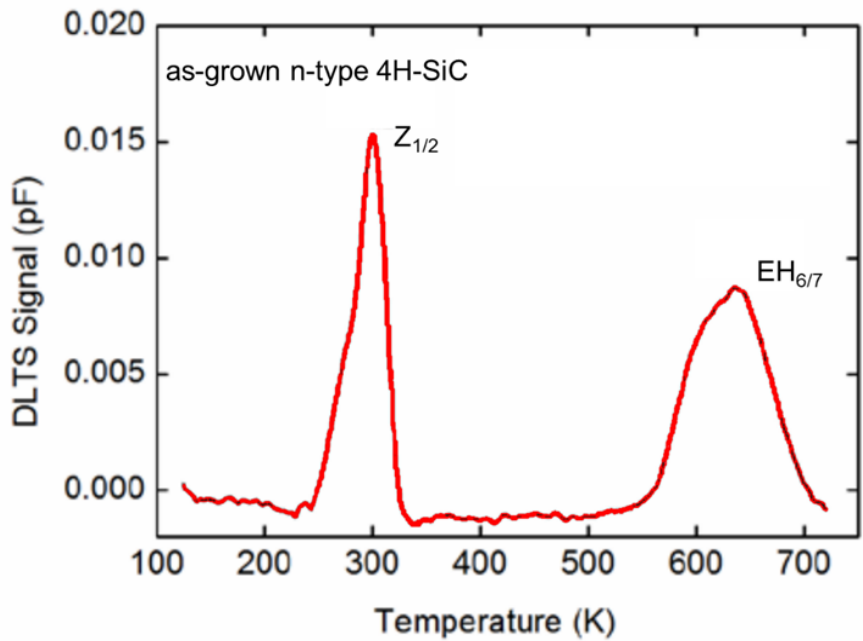

**Figure 4.** DLTS spectrum for as-grown n-type 4H-SiC. Data taken from Ref. [40].

$Z_{1/2}$ and $EH_{6/7}$ are the most common traps in 4H-SiC, introduced either during the crystal growth or upon irradiation. The $Z_{1/2}$ concentration can be further increased by irradiation or high-temperature annealing, which reduces carrier lifetime. This is the reason why $Z_{1/2}$ is often called a "lifetime killer" within the SiC community [41,42,43]. However, numerous studies have shown that thermal oxidation [44] or carbon implantation followed by subsequent annealing [45,46,47] could reduce the concentration of $Z_{1/2}$ and $EH_{6/7}$ traps in 4H-SiC.

As shown in Figure 4, the $Z_{1/2}$ introduces an asymmetric peak in the DLTS spectrum. In their pioneering work on $Z_{1/2}$ using DLTS, Hemmingsson et al. [48] showed that $Z_{1/2}$ is the superposition of two nearly identical $Z_1$ and $Z_2$ negative-U defect transitions, each located on a different sub-lattice site. Later on, by connecting electron paramagnetic resonance (EPR) and photo-EPR data with the DLTS results, it was possible to ascribe $Z_{1/2}$ to transitions involving the carbon vacancy ($V_C$) in 4H-SiC on distinct sub-lattice sites and to assign $Z_{1/2}$ to a transition between double negative and neutral charge state of carbon vacancy $V_C$ (=/0) [49]. Additionally, studies using the L-DLTS technique have provided direct evidence that the $Z_{1/2}$ comprises two distinct components, namely $Z_1$ and $Z_2$. These components were identified with activation energies for electron emission of 0.59 eV and 0.67 eV, respectively, and were attributed to the (=/0) charge transitions of $V_C$ located at $-h$ and $-k$ lattice sites in the 4H-SiC crystal [50,51]. Figure 5 shows L-DLTS spectra for $Z_1$(=/0) and $Z_2$(=/0).

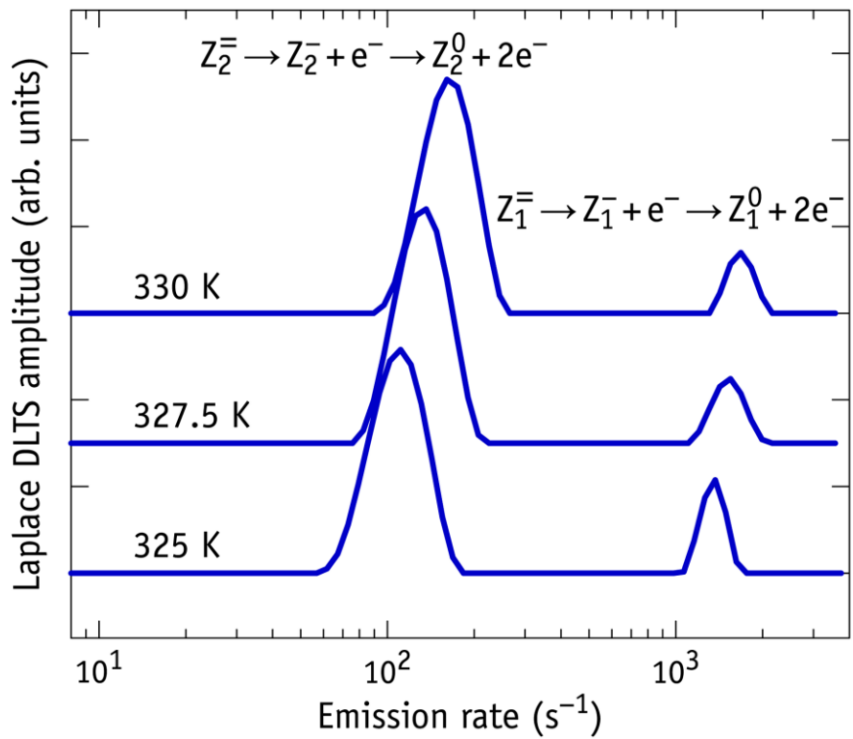

**Figure 5.** L-DLTS spectra for $Z_1$ and $Z_2$ observed in as-frown n-type 4H-SiC sample. Reprinted from Ref. [51], with the permission of AIP Publishing.

Moreover, significant efforts were also undertaken to resolve the broad $EH_{6/7}$ peak (Figure 4). Danno et al. [52] have successfully distinguished $EH_6$ and $EH_7$ through simulations of the Fourier-transform DLTS spectrum of the $EH_{6/7}$ peak. Later, direct evidence that $EH_{6/7}$ comprises two distinct components was provided by Alfieri et al. [53].

They have identified two separate energy levels, 1.30 eV and 1.49 eV, corresponding to EH$_6$ and EH$_7$, respectively.

2.4 Carbon antisite – carbon vacancy (C$_{Si}$-V$_C$) pair

The traps labeled as EH$_4$ and EH$_5$ are commonly observed in DLTS spectra of ion-implanted or irradiated n-type 4H-SiC samples [54,55,56]. The EH$_4$ peak reaches its maximum at approximately 400 K (as shown in Figure 6), while the EH5 peak appears at slightly higher temperatures (not shown here). The reported activation energies for these defects are approximately 1.0–1.1 eV below the conduction band. Also, it is a practice to refer to these traps as EH$_{4/5}$.

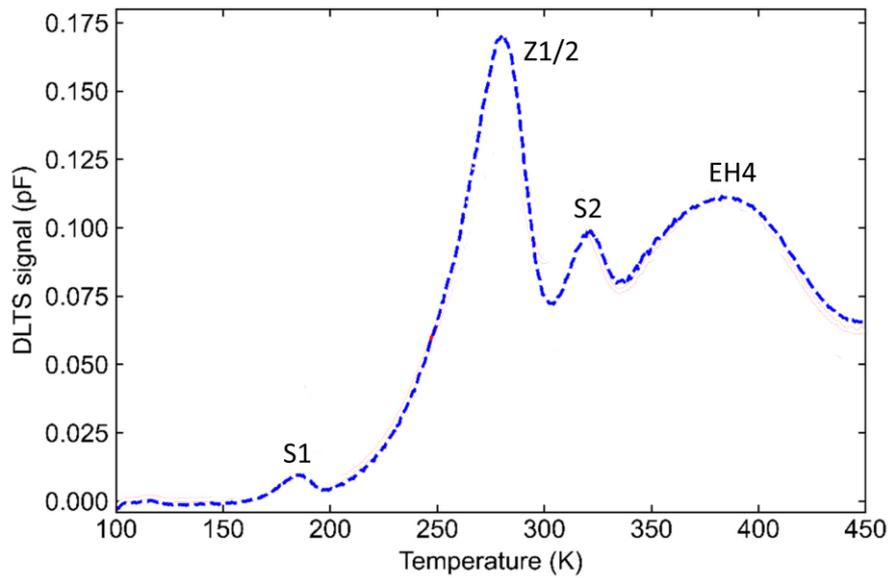

**Figure 6**. DLTS spectrum of neutron-irradiated n-type 4H-SiC sample. Data modified from Ref. [54].

Karsthof et al. [55] have investigated the EH$_4$ and EH$_5$ traps using DLTS and photoluminescence (PL) in proton-irradiated and subsequently annealed n-type 4H-SiC samples. In their study, they have proposed that EH$_4$ (E$_C$ - 1.0 eV) and EH$_5$ (E$_C$ - 1.1 eV) arise from the (+/0) charge transition level associated with different configurations of the carbon antisite-carbon vacancy (C$_{Si}$-V$_C$) pair.

This conclusion was further validated by Nakane et al. [56], who utilized 2 MeV electron irradiation followed by annealing to introduce deep-level defects, which were analyzed using photo-induced current transient spectroscopy (PICTS) and EPR. Their study identified several deep-level defects, including two with energies of approximately 0.72 eV and 1.07 eV, which were attributed to the (0/−) and (+/0) charge transitions of the C$_{Si}$-V$_C$ pair, respectively.

2.5 Boron-related defects

Alongside aluminum, boron is among the most widely used p-type dopants in SiC. However, its unintentional incorporation during crystal growth has been frequently reported. This phenomenon is attributed to the presence of boron impurities in the graphite susceptor used during the chemical vapor deposition (CVD) process [57,58,59,60].

Storasta et al. [61] investigated this issue by employing MCTS to examine the effects of unintentional boron incorporation in low-doped n-type 4H-SiC. The SiC samples were grown using a hot-wall CVD process. Their findings confirmed earlier reports that boron introduces two distinct deep levels, referred to as shallow boron (B) and deep boron (D-center). Figure 7 shows MCTS spectra for as-grown and neutron irradiated n-type 4H-SiC samples: (a) shallow boron, B and (b) deep boron, D-center. The activation energies for hole emissions for B and D-center are estimated as $E_v + 0.28$ eV and $E_v + 0.61$ eV, respectively [33, 61, 62, 63, 64].

Bockstedte et al. [64] further identified the origin of these boron-related levels. The shallow boron level (B) was attributed to boron atoms substituting silicon atoms ($B_{Si}$) in an off-center configuration. The D-center, associated with deep boron, was assigned to boron atoms substituting carbon atoms ($B_C$) in a perfect substitutional configuration. This defect was initially observed to create a deep level at ~0.60 eV above the valence band based on DLTS measurements. However, recent L-DLTS studies revealed that the D-center consists of two emission lines with activation energies of $E_v+0.49$ eV and $E_v+0.57$ eV [33]. These levels are associated with boron atoms occupying -$h$ and -$k$ lattice sites, respectively. Although $V_C$ has been identified as the dominant carrier lifetime limiting defect in 4H-SiC, Ghezellou et al. [63] have reported an extensive study on boron-related defects being another dominant source of recombination and acting as lifetime limiting defects in 4H–SiC material.

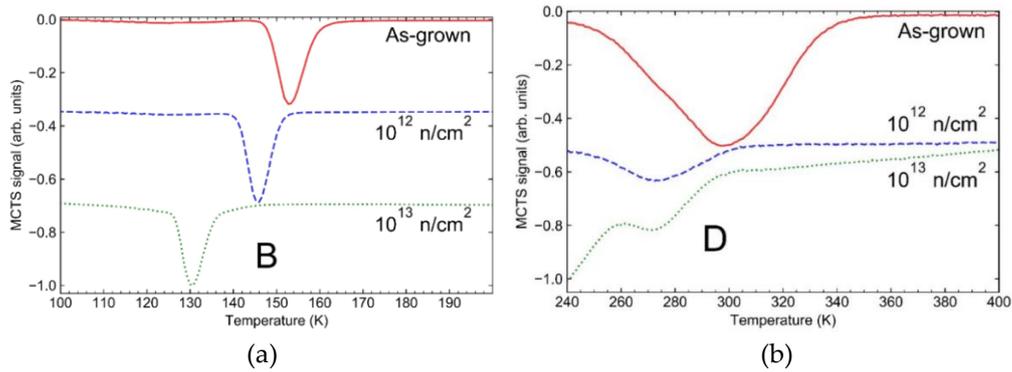

**Figure 7.** MCTS spectra for as-grown and neutron-irradiated n-type 4H-SiC samples. (a) shallow boron, B and (b) deep boron, D-centre. Reprinted from Ref. [33] with permission from Elsevier.

In this section, we have provided an overview of the most significant electrically active defects in 4H-SiC. These defects can either be intrinsic, present in as-grown material, or induced by various radiation sources, such as electrons, protons, ions, and neutrons. Table 2 summarizes the information on the most important defects in 4H-SiC.
.

Table 2. Details on the most common defects in 4H-SiC. Activation energies are estimated from DLTS, L-DLTS, MCTS and L-MCTS measurements.

| Trap Label | Identification | Activation energy (eV) | References |
|---|---|---|---|
| $EH_1$ | $C_i$ | $E_c - 0.40$ | [34,35,36] |
| $EH_3$ | $C_i$ | $E_c - 0.70$ | [34,35,36] |
| $S_1$ | $V_{Si}$ (-3/-) | $E_c - 0.40$ | [36,37,38] |
| $S_2$ | $V_{Si}$ (=/-) | $E_c - 0.70$ | [36,37,38] |
| $Z_1$ | $V_c$ (=/0) | $E_c - 0.59$ | [48,49,50,51] |
| $Z_2$ | $V_c$ (=/0) | $E_c - 0.67$ | [48,49,50,51] |
| $EH_4$ | $C_{Si}-V_C$ | $E_c - 1.00$ | [55,56] |
| $EH_5$ | $C_{Si}-V_C$ | $E_c - 1.10$ | [55,56] |
| $EH_6$ | $V_c$ (0/++) | $E_c - 1.30$ | [52,53] |
| $EH_7$ | $V_C$ (0/++) | $E_c - 1.40$ | [52,53] |
| B | $B_{Si}$ | $E_v + 0.28$ | [33,61,62,63] |
| D-center | $B_C$ | $E_v + 0.61$ | [33,61,62,63] |

It is important to note that the electrically active defects discussed here do not represent an exhaustive list. Other defects, including those associated with impurities such as nitrogen [65] and vanadium [66], have also been observed in 4H-SiC. However, we have deliberately chosen not to include them in this discussion. The primary objective of this review is to maintain a focused and structured approach, making it easier to follow, rather than attempting to provide an exhaustive survey of all reported defects in the literature.

By narrowing our focus, we aim to highlight the most relevant defects in the context of their impact on the electrical properties of 4H-SiC, particularly those that are crucial for its applications in power electronics, quantum technologies, and radiation detection. Future studies may expand upon this foundation by incorporating a broader range of defect types, including impurity-related and process-induced defects.

## 3. 6H-SiC

6H-SiC is distinguished by its wide bandgap (3.02 eV), high thermal conductivity, and exceptional robustness under extreme conditions, making it a highly versatile material for applications in high-temperature electronics, radiation-hardened devices, and optoelectronics. However, despite these advantages, its lower electron mobility compared to 4H-SiC (as shown in Table 1) limits its suitability for power electronic applications, where high carrier mobility is essential for achieving efficient performance. This limitation has resulted in a significantly lower number of published studies on intrinsic or radiation-induced defects in 6H-SiC compared to 4H-SiC, which remains the preferred polytype for most power and high-frequency applications. In this section, we provide an overview of the most significant achievements in the study of electrically active defects in 6H-SiC.

3.1 Carbon vacancy ($V_C$)

Figure 8 shows DLTS spectra of as-irradiated n-type 6H-SiC. The low-energy electron irradiation (E=150 keV) with fluencies $1 \times 10^{16}$, $2 \times 10^{16}$ and $4 \times 10^{17}$ cm$^{-2}$ was used [67]. It should be noted, that $E_{1/2}$ (0.45 eV) and R (1.25 eV) traps were already present in the DLTS spectrum for as-grown 6H-SiC materials, while $RD_5$ (0.57eV) and ES (0.80eV) traps are observed only after low-energy electron irradiation.

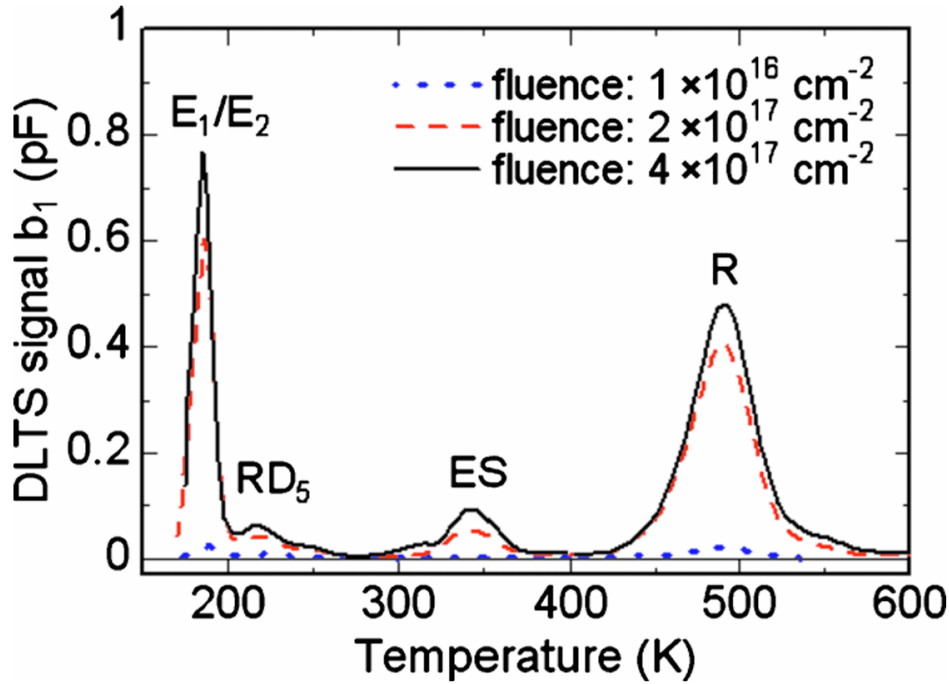

**Figure 8**. DLTS spectra of low-energy electron irradiated n-type 6H-SiC samples for several fluencies. Reprinted from Ref. [67], with the permission of AIP Publishing.

Based on defect generation rates and annealing behavior, Sasaki et al. [67] proposed that $E_{1/2}$ and R in n-type 6H-SiC correspond to $Z_{1/2}$ and $EH_{6/7}$ in n-type 4H-SiC (Section 2) and are attributed to the carbon vacancy ($V_C$). Similar to the $Z_{1/2}$ in 4H-SiC, the $E_{1/2}$ trap in 6H-SiC has attracted significant research attention. Initially, the $E_{1/2}$ DLTS peak was separated into two components, leading to the interpretation that $E_1$ and $E_2$ correspond to $V_C$ at two distinct lattice sites, -h and -k. However, this interpretation was challenged by Pensl et al. [68], who correctly pointed out that 6H-SiC has three non-equivalent lattice sites for vacancies: one hexagonal (-h) and two cubic (-$k_1$ and -$k_2$). Their analysis suggested that a three-component structure was more likely, rather than just two. Building upon this hypothesis, Koizumi et al. [69] successfully resolved the $E_{1/2}$ into three distinct emission components using L-DLTS. As illustrated in Figure 9, their results confirm that $E_{1/2}$ consists of three individual peaks (labeled as $E_{2L}$, $E_{2H}$ and $E_1$), corresponding to the acceptor levels of carbon vacancies at the three non-equivalent lattice sites in 6H-SiC. This finding represents a major advancement in the understanding of $V_C$ defects in 6H-SiC, offering greater insight into their electronic properties.

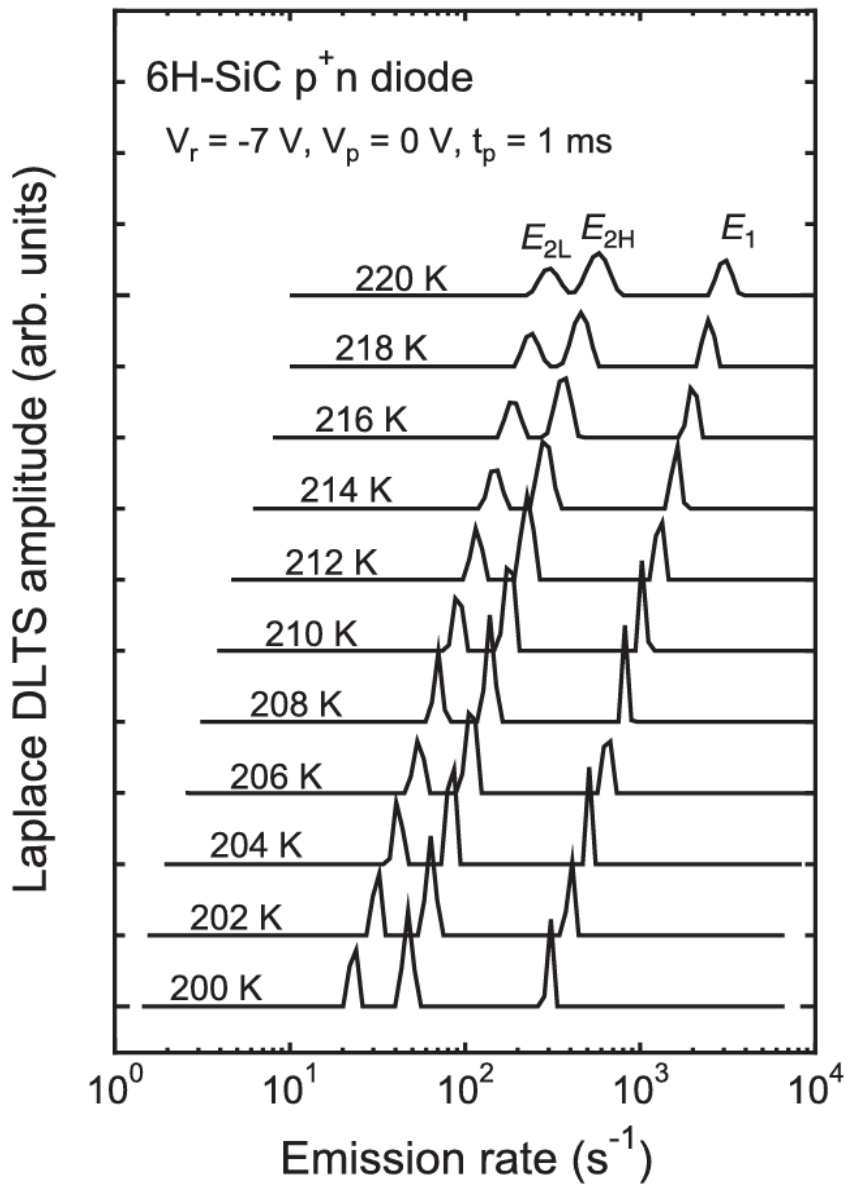

**Figure 9.** L-DLTS spectra for 6H-SiC p+n diode. Measurements are done at different temperatures, from 200 up to 220 K. Reprinted from Ref. [69], with the permission of AIP Publishing.

3.2  Carbon interstitials ($C_i$)

As already mentioned, $RD_5$ and ES traps (shown in Figure 8) are not observed in as-grown 6H-SiC material but they only appear upon irradiation. Sasaki et al. [67] investigated the annealing behavior of these defects, and their findings indicate that both $RD_5$ and ES are thermally unstable. Interestingly, the similarities between the $RD_5$ and ES traps in n-type 6H-SiC and the $EH_1$ and $EH_3$ traps in n-type 4H-SiC are striking, naturally leading to the assumption that they originate from the same underlying defects. These similarities include: (i) the fact that all these traps are induced by low-energy electron irradiation (E<200keV), and (ii) they anneal out at relatively low temperatures. Based on experimental and theoretical evidence, Sasaki et al. [67] have proposed that $RD_5$ and ES are associated with carbon interstitials ($C_i$).

Additionally, a broad peak appearing at a similar temperature position as ES has been observed and labeled as $Z_{1/2}$ [70]. This designation can be misleading, as $Z_{1/2}$ in 6H-SiC has no connection whatsoever with $Z_{1/2}$ in 4H-SiC. The $Z_{1/2}$ trap is observed in n-type 6H-SiC upon 2.5 MeV He+ ion implantation [70]. The activation energy for electron emission of $Z_{1/2}$ (0.71eV) is close to the activation energy of ES (0.80eV), but there is no conclusive evidence proving that these traps originate from the same defect. Ruggiero et al. [70] attributed the $Z_{1/2}$ trap to the divacancy ($V_C$-$V_{Si}$), based on DLTS and photoluminescence (PL) measurements. Applying a comparative approach with 4H-SiC, where traps ($S_{1/2}$) are introduced by high-energy particles such as protons or ions have been identified as silicon vacancy ($V_{Si}$), it is reasonable to assume that the $Z_{1/2}$ trap is indeed related to divacancies rather than carbon interstitials.

**Table 3.** Electrically active defects in 6H-SiC. Activation energies are estimated from DLTS measurements.

| Trap Label | Identification | Activation energy (eV) | References |
|---|---|---|---|
| $E_{1/2}$ | $V_C$ | Ec - 0.45 | [67, 68, 69] |
| $RD_5$ | $C_i$ | Ec - 0.57 | [67, 69] |
| $Z_{1/2}$ | $V_C$ - $V_{Si}$ | Ec - 0.71 | [69] |
| ES | $C_i$ | Ec - 0.80 | [67] |
| R | $V_c$ | Ec – 1.25 | [67,69] |

3.3 Applications

As discussed in Section 2, point defects such as the silicon vacancy ($V_{Si}$) have shown great promise for quantum technology applications and have been extensively studied in 4H-SiC [5,6,7,37,39]. These defects exhibit long spin coherence times and optically addressable spin states, making them highly relevant for quantum computing, sensing, and communication.

However, the physics of similar defects in other SiC polytypes remains significantly less explored. Understanding their properties across different polytypes is crucial, as variations in stacking order and crystal symmetry can strongly influence defect formation, charge states, and optical transitions.

Recently, Ousdai et al. [71] investigated the formation and thermal stability of point defects in 6H-SiC using PL spectroscopy. Their findings are promising, revealing defect-related emissions that suggest 6H-SiC could also be a viable platform for quantum applications. These results open up a new research direction, highlighting the potential of 6H-SiC for quantum technologies and warranting further studies to explore its suitability in areas such as spin qubits and single-photon emission.

Future investigations will be essential to fully characterize the quantum coherence properties of defects in 6H-SiC and compare their performance with those in 4H-SiC. If successful, this could broaden the scope of SiC-based quantum devices and expand the material choices available for next-generation quantum technologies.

It is worth mentioning that 6H-SiC has already found its place in several niche applications. Thanks to its high crystalline quality, 6H-SiC is used as a substrate for growing epitaxial layers of 4H-SiC or GaN [72,73]. Additionally, 6H-SiC polytype has been proposed for UV and visible light devices [74].

## 4. 3C-SiC

For many years, the 3C-SiC polytype has been proposed as the most promising SiC polytype for electronic device applications due to its unique advantages. These advantages include lower production costs than 4H-SiC and most importantly its ability to be grown epitaxially on large-area, cost-effective silicon (Si) wafers. This compatibility with existing Si-based semiconductor technology facilitates easier integration into established manufacturing processes [75]. However, this compatibility did not come without the price. The mismatch in the lattice constant and thermal expansion coefficient between 3C-SiC and Si can introduce electrically active defects [76]. Moreover, 3C-SiC possesses a lower bandgap of 2.36 eV compared to 4H-SiC (3.26 eV) and 6H-SiC (3.02 eV). This fact limits its performance in high-power and high-temperature devices where wider bandgaps are advantageous.

### 4.1 Electrically active defects

The number of available studies on 3C-SiC is significantly lower compared to its 4H- and 6H-SiC counterparts (Figure 1). In the early investigations of defects in 3C-SiC, a variety of deep-level defects were reported. However, it was initially unclear whether these defects originated from the growth process, were related to the Si/3C-SiC interface, or were intrinsic to 3C-SiC itself. A pioneering effort in this field was led by M. Kato [76, 77,78], who systematically examined defects in 3C-SiC and identified several traps arising from the 3C-SiC/Si interface. This work provided crucial insights into distinguishing interface-related states from bulk defects in 3C-SiC. Later studies expanded on this foundation by investigating both as-grown and radiation-induced defects in 3C-SiC.

Nagesh et al. [75] investigated neutron-irradiated n-type and p-type 3C-SiC. In n-type material, they observed an electron trap located at $E_C-0.49$ eV, while in p-type material, they identified three hole traps at $E_V + 0.18$ eV, $E_V + 0.24$ eV, and $E_V + 0.51$ eV. These findings highlight the impact of neutron irradiation on carrier trapping and recombination processes in 3C-SiC. Weidner et al. [79] investigated intrinsic defects introduced in n-type 3C-SiC by $H^+$ and He implantation, as well as high-energy electron irradiation. They reported the presence of nine distinct deep-level defects, labeled $W_1$-$W_9$. Among these, the W6 trap, with an activation energy for electron emission of 0.58 eV, was the most dominant. Remarkably, $W_6$ showed strong similarities to the $Z_{1/2}$ trap in 4H-SiC, which has been attributed to the carbon vacancy ($V_C$). Further investigations by Beyer et al. [80] provided additional confirmation that the $E_1$ trap (0.57 eV) observed in their study of n-type 3C-SiC corresponds to the $W_6$ trap described by Weidner et al. [79], and they assigned it to the carbon vacancy in 3C-SiC. This identification is a crucial step toward understanding the role of vacancy-related defects in 3C-SiC, as $V_C$ plays a pivotal role in carrier recombination. The carbon vacancy's significance extends across all SiC polytypes, and it is the only defect that has been extensively studied and convincingly confirmed in 3C, 4H, and 6H-SiC polytypes.

### 4.2 Applications

With its cubic structure and inherent compatibility with existing silicon technology, 3C-SiC presents unique opportunities for niche applications that leverage its exceptional material properties. While it has not yet reached the same level of industrial maturity as 4H-SiC, ongoing research continues to highlight its potential in several key areas. These applications include, but are not limited to:

(i) Solar Cells:

3C-SiC's excellent thermal stability and a bandgap wider than Si (1.12eV) make it a strong candidate for next-generation solar cells. Unlike traditional silicon-based solar cells,

3C-SiC can operate efficiently in extreme environments, including high radiation and high temperatures [81,82,83].

(ii)   Quantum applications:

Similar to its 4H and 6H counterparts, the potential of 3C-SiC for quantum qubits and quantum sensing has gained attention [84,85,86]. However, as it is the case with other SiC polytypes, the primary challenge remains defect engineering. This involves the precise introduction and control of individual defect centers, which is crucial for harnessing the material's full capabilities in quantum applications.

(iii)   Biomedical Devices and Bioelectronics:

One of the most promising application areas for 3C-SiC is for biomedical devices and bioelectronics [87]. 3C-SiC is recognized as a bio- and hemo-compatible material, meaning it does not induce immune responses or blood clotting upon contact with biological tissues [88]. These properties make it highly suitable for implantable medical devices and biosensors [89,90].

In summary, the characterization of electrically active defects in 3C-SiC remains an open field of research. Compared to 4H-SiC, where defect identification has reached a mature stage (at least for n-type material), many questions regarding the nature, stability, and impact of electrically active defects in 3C-SiC are still open. Further investigations combining DLTS with complementary spectroscopic techniques such as EPR and PL are necessary to provide a more comprehensive understanding of defect physics in 3C-SiC and its impact on electronic and optoelectronic device applications.

## 5. Conclusion

Figure 10 shows a diagram with the most common electrically active defects in 3C-SiC, 4H-SiC, and 6H-SiC. As highlighted from the outset, and further emphasized in Figure 1, there is a significant discrepancy in the number of published studies focusing on these polytypes. Specifically, while defect-related research on 4H-SiC has been extensive and comprehensive over the years, studies on 3C-SiC and 6H-SiC remain relatively limited.

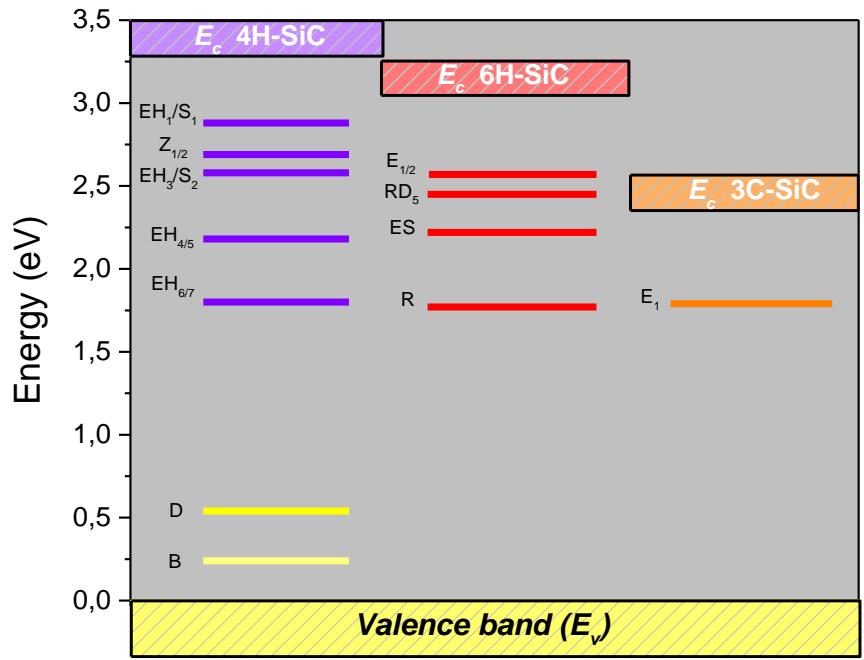

**Figure 10.** Graphical representation of the most common energy levels in the band gaps of 3C, 4H and 6H-SiC based on DLTS measurements.

Due to the sheer volume of research, we can confidently state that most aspects of the defect landscape in 4H-SiC have been well characterized, with many pieces of the puzzle now fitting into place. In contrast, the understanding of defects in 3C-SiC and 6H-SiC is still evolving, with significant gaps that require further investigation. Moreover, it is evident that 4H-SiC has become the dominant polytype in power electronics, where it serves as the key material driving advancements in high-performance semiconductor devices. Additionally, 4H-SiC is steadily gaining recognition as a promising candidate for emerging applications in quantum technologies and radiation detection. On the other hand, 3C-SiC and 6H-SiC have not yet reached the same level of technological maturity, limiting the full exploration of their potential applications. In this review paper, we have addressed specific aspects of defect-related challenges in 3C-SiC and 6H-SiC, shedding light on niche areas that warrant further attention. However, as research efforts on these polytypes continue to expand in the future, the scope of investigation will undoubtedly broaden, revealing new applications.